\documentclass[3p,times]{elsarticle}
\usepackage{amssymb}
\usepackage{amsbsy}
\usepackage{bm}
\usepackage{latexsym}
\usepackage[active]{srcltx}
\usepackage{latexsym,amsfonts,amsthm,amsmath,amscd,amssymb,color}
\usepackage{natbib}

\renewcommand{\theequation}{\arabic{equation}}
\numberwithin{equation}{section}

\def\spazio#1{\vrule height#1em width0em depth#1em}
\biboptions{sort&compress}
\begin{document}
%
%
	\title{\bf {Relativistic levels of mesic atoms}}
	\author{{R. Giachetti}}
	\address{{Physics Department, Universit\`a di Firenze, Italy\\
			Istituto Nazionale di Fisica Nucleare, Sezione di Firenze, Italy}}
	\author{{E. Sorace}}
	\address{{Istituto Nazionale di Fisica Nucleare, Sezione di Firenze, Italy}}
%
%
\begin{abstract}
{We revisit the derivation of
the covariant two-body scalar-fermion equation with
a Coulomb interaction,
presented in a previous paper. We show that it
can be given the formal aspect of a Dirac equation, but for the fact that the eigenvalue is also contained
in one of the coefficients and thus it is not
linearly included. The discussion of the boundary
value problem is therefore different, although some
properties of the Dirac equation can be recovered in
an approximation bringing back to the 
concept of reduced mass.
We discuss a mixed analytic-numerical
method of solution which allows to obtain very accurate results and we calculate the lowest levels, states and QED corrections for pionic
and kaonic atoms.  }


\end{abstract}
\maketitle
\medskip

\section{Introduction}
\medskip

Some years ago we began a research program on the two body
covariant wave equations and we developed a   relativistic
wave equation for two interacting fermions 
\cite{GS1,GS2,GS3,BGS_JPB,BGS_PR}. We studied the
covariance, the different
limits and the spectral properties of such an equation and
the first applications were made on the hyperfine
levels of the Hydrogenic atoms. We then addressed to high
energy physics, calculating the mass of the mesons and
their radiative decays. 
In all cases the results were in excellent
agreement with experimental results, even when dealing with
the masses of light
mesons for which potential models had generally failed. 
The essential reason for the
accuracy of the results can be traced back to the complete
covariance of the
treatment,  which includes all the relativistic and
the two-body effects.
In more recent times, in order to broaden the applicability of
the method, we have extended our analysis by
considering relativistic  objects of 
different nature. We have thus determined the 
wave equations for interacting scalar-scalar and
scalar-fermion particles
\cite{GS-AP}, enabling thus the study of further elementary two-body systems.
Again we have proved the covariance, the different
limits of such equations and the general properties
of their spectrum, giving them a clear physical interpretation. 

In this paper we revisit and we give a further
development of the relativistic scalar-fermion
equation together with its application  to the 
pionic and kaonic atoms. 
The physical interest for such systems is due to
the fact that meson masses are two-three order of
magnitude larger than the
electron mass, their atomic orbits are of the order of hundred fm 
and their binding energies of some
keV. Therefore they provide a
particularly useful opportunity for studying nuclear
interactions in QCD low energy regime \cite{Iwa,Schroeder,Curc}. 
The effects of the strong interactions on the
spectrum is given by a shift of the purely
electromagnetic levels and an absorption width,
both of them experimentally 
measured through X-ray spectroscopy \cite{,Bazzi}. 
In fact the size of the atom is much larger
than the range of strong interactions, so that QED
itself provides almost exact values of the atomic
levels and the fundamental state is actually the only
one for which the nuclear force plays a role. Hence
the difference between the
measured transition energies from excited states to
the fundamental one and
the corresponding energies calculated by pure 
QED are essentially the same as the
shift of the fundamental level. A history of the
determination of the pion-nucleon coupling constant
can be found in \cite{Matsinos}.
Up to the most recent papers \cite{IndeTra, Tra}, 
the mass of the $\pi$ meson has been evaluated by
matching the measured values of the  photon energy
emitted in the transitions between two rather
high level of  pionic atoms.
The energy of those same decays were calculated by
the  Schr\"odinger equation with a perturbation 
treatment of the relativistic and QED effects \cite{SBIP}.  
Different corrections of a more phenomenological
nature have also been considered, such as, for
instance, the finite dimensions of the particles,
usually treated as if they were uniformly charged
spheres with a non vanishing radius
\cite{Guy,Pato,XGG}. 
A different measure involving the $\mu$-neutrino
mass has been presented in \cite{DFK}.

We aim here at a more essential treatment which maintains 
the covariance, such to take exactly 
into account the relativistic effects. 
We therefore will add the lowest order QED
corrections, due to vacuum polarization, while assuming a 
basic model formed by two particles which are
point-like and interact by  a Coulomb potential.
Of course this does not mean that we believe that
phenomenological corrections are negligible, 
but only that we defer their inclusion until we find
a formulation fitting the covariant two body framework.
For instance, we know that it is comparatively easy
to introduce a term mimicking a finite radius of 
an attracting center for the case of a single particle in an external 
potential. For two interacting points, after the separation of the
global coordinate, the relative part becomes an
entangled system and the definition of a potential producing 
a finite radius effect in terms of the 
relative coordinate only seems tough. Starting
with extended particles would directly lead to the
formulation of a field theory producing the appropriate 
form factors. This is outside of our
present purposes.  
On the other hand we notice that the final
reduced form of the scalar-fermion equation develops
a radial term clearly related to relativistic 
corrections of the the classical reduced mass. 
Thus, besides the results, we also take
the opportunity to illustrate some properties of the
fermion-scalar equation, as well as to explain the 
ideas of the mixed analytical-numerical treatment 
we have used in order to obtain a very
accurate solution of the spectral problem. 
High precision appears to be necessary, as seen by
the numerical results, since the spectrum presents
pairs of very close levels, whose difference becomes 
fainter and fainter with increasing energy. 
These states appear as a direct consequence of the
covariant treatment which takes into account the
fermionic nature of the proton. 
For the sake of completeness, we begin Section 2 with
a sketchy derivation of the scalar-fermion equation
relegating to Appendix I the summary
of the kinematical variables introduced in our
previous papers. 
We then show that the scalar-fermion equation can be 
cast into the the form of
a Dirac equation in a central potential,
but for the fact that
the reduction procedure makes a coefficient depending 
upon the eigenvalue. Unfortunately
this non-linear inclusion of the eigenvalue
forbids the application of most results holding 
for the Dirac equation \cite{BLP} and produces,
in particular, the close pairs of levels we mentioned above.
In Section 3 some  details of the discussion 
of the boundary value problem are given. In
particular we work out the series solutions and
the recurrence relations at the two singular points,
the origin and infinity, as well as at any regular
point of the positive real line. 
We then present the results for the pure Coulomb
relativistic levels of the pionic and kaonic atoms.
In Section 4 we state the form of the
potentials we have used for the one- and two-loop 
vacuum polarization potentials. The detailed results,
obtained with the use of our eigenfunctions, are presented 
in the table given in Appendix II. It can be seen that 
the one-loop and irreducible two-loop corrections are very regular,
namely  monotonically decreasing for increasing
energy of the levels. The apparently more bizarre behavior 
of the corrections due to reducible two-loop graph 
comes from a change of sign of 
the corresponding potential. 
As a consequence,  the overall results are very sensitive
to the wave-function density distribution and 
a little bit longer to be calculated with 
the desired accuracy.
In particular, for some levels,  the different 
ratios of the meson to proton mass of
pionic and kaonic atoms may produce effects with opposite sign.  
In any case these last corrections are one-two orders less than
the previous ones. A brief comment is also in order about the
seven decimal figures which we have presented 
the results with. 
All of them are numerically meaningful, although supernumerary
for a comparison with the experimental results. Their inclusion 
is motivated in order to give a clear distinction of the levels 
and to show the regular trend of the behavior both of spectrum and corrections.
Some brief conclusions are given in the last Section 5.  
In the final Appendix III we show an intriguing
feature of our scalar-fermion equation, in which it
is possible to single out an interaction term
related to the corrections connected with the classical 
'reduced mass'. Indeed, although not directly 
deduced from the covariant framework, this quantity
gives an approximation of the levels
to the first order of the fine structure constant.
Relativistic corrections to the reduced mass improve
the accuracy of the levels \cite{MF}.

\medskip
 
%
\section{The scalar-fermion equation and its spectrum}
%
\bigskip

In this section we recall the basic steps leading 
to the relativistic wave equation for a fermion
and a scalar particle interacting through a Coulomb potential \cite{GS-AP}. 
We then make explicit the dimensionless form of the 
second order system which defines the
boundary value problem leading to the spectrum.
An improvement with respect to the original derivation is
the Dirac form the reduced equation can be given, which
makes somewhat more familiar the treatment. 
The solution has been discussed analytically
as far as possible in order to obtain a high precision 
for the spectral levels. In the next Section we will therefore   
develop the recurrence relations which make it 
possible a rapid achievement of very accurate numerical results. 
As we said in the
Introduction, the physical systems we have considered are the pionic atom $\,({H}^+\!,\,\pi^-)\,$ and the kaonic atom 
 $\,({H}^+\!,\,{K}^-),$ whose lowest Coulomb levels are given in Tables 1 and 2.

Let us recall some elementary properties of the 
Klein-Gordon (KG) and Dirac equations
in order to establish the two-body relativistic 
equation for a scalar and a fermion. The covariance of the
method has been largely discussed in the series 
of our previous papers 
\cite{GS1,GS2,GS3,BGS_JPB,BGS_PR,GS-AP}, so that, 
for the sake of completeness,
we just report in Appendix I the definition of 
the phase-space we are using and some few words to 
describe its physical and geometrical properties.

A Hamiltonian formulation for the Klein-Gordon 
equation  can be given in the form of a 2-dim system: 
\begin{eqnarray}
i\,{\partial\Phi}/{\partial t}-H_S\,\Phi=0\,,
\quad \Phi = {}^{T}\bigl(\phi_1,\phi_2\bigr)
\label{KG_equation_FV}
\end{eqnarray}
In the standard notation $\,\sigma_z,\,\sigma_\pm\,$  for the Pauli matrices, the Hamiltonian $H_S$ has the expression 
\begin{eqnarray}
H_S = -\,\sigma_-\,\,\bigl(\,{{\nabla}^2}/{2m_S}\,\bigr) +(\sigma_z+2\sigma_+)\,m_S
=\left(\begin{matrix}
m_S & 2m_S \spazio{0.2}\cr
-{{\nabla}^2}/{2m_S} & -m_S 
\end{matrix}
\,\right)
\label{HS}
\end{eqnarray}
where $\,m_S\,$ is the mass of the scalar.
It is obtained from the Feshbach-Villars \cite{FV} 
representation of the KG equation with a transformation generated by
$\sigma_+ + \sigma_z$.

The free Dirac Hamiltonian $H_F$ in spherical coordinates reads
\begin{eqnarray}
H_F = \left(\begin{matrix}
m_F & 0 & q_0 & \sqrt{2}\,q_- \spazio{0.2}\\
0 & m_F & -\sqrt{2}\,q_+ & q_0 \spazio{0.2}\\
q_0 & \sqrt{2}\,q_- &  -m_F & 0 \spazio{0.2}\\
-\sqrt{2}\,q_+ & -q_0 & 0 & -m_F 
\end{matrix}\right)
\label{HF}
\end{eqnarray}
where $\,m_F\,$ is the mass of the fermion and the spherical derivatives $q_+$, $q_-$ and $q_0$ are defined by
\begin{eqnarray}
q_\pm=-( \, \pm{q_{x}} + i\,{q
	_{y}}\,)/\sqrt{2}\,,\qquad\qquad q_0={q_{z}}\,,\qquad\qquad \,q_k \rightarrow 
-i\partial/\partial r_k\,
\label{Xpm0}
\end{eqnarray}
Denoting the fermion radial coordinate by $\xi=(x^2+y^2+z^2)^{1/2}$, the use of the spherical spinors
\begin{equation}
\Omega_{\,\ell+\frac12,\,\ell,\,m}(\theta,\phi) =
\begin{pmatrix}
\bigl({j+m}\bigr)^{1/2}\,\bigl({2j}\bigr)^{-1/2}\,Y_{\ell,\,m-\frac12}(\theta,\phi)
\spazio{0.4}\\
\bigl({j-m}\bigr)^{1/2}\,\bigl({2j}\bigr)^{-1/2}\,Y_{\ell,\,m+\frac12}(\theta,\phi)
\end{pmatrix},
~~
\Omega_{\,\ell-\frac12,\,\ell,\,m}(\theta,\phi) =
\begin{pmatrix}
-\bigl({j-m+1}\bigr)^{1/2}\,\bigl({2j+2}\bigr)^{-1/2}\,Y_{\ell,\,m-\frac12}(\theta,\phi)
\spazio{0.4}\\
\phantom{-}\bigl({j+m+1}\bigr)^{1/2}\,\bigl({2j+2}\bigr)^{-1/2}\,Y_{\ell,\,m+\frac12}(\theta,\phi)
\end{pmatrix}
\end{equation}
allows for the construction of a Dirac state where 
angular momentum and parity are diagonal:
\begin{eqnarray}
\psi(\xi,\theta,\phi)=
\begin{pmatrix}
\varphi(\xi,\theta,\phi)\spazio{0.4}\\ 
\chi(\xi,\theta,\phi)
\end{pmatrix}
= 
\begin{pmatrix}
a(r)\,\,\Omega_{j\ell m}(\theta,\phi)\spazio{0.4}\\
b(r)\,\,\Omega_{j\ell' m}(\theta,\phi)
\end{pmatrix}\,,\qquad \ell=j-\frac12,\quad \ell'=j+\frac12
\label{Spinore_di_Dirac}
\end{eqnarray}
We now write the relativistic Hamiltonian for a scalar and a fermion interacting through a Coulomb potential 
expressed in the coordinates introduced in Appendix I:
\begin{eqnarray}
H=H_S\otimes\mathbf{I}_{4}+\mathbf{I}_{2}\otimes H_F + \mathbf{I}_{4}\,V(r)\,,\qquad V(r)=-\alpha/r 
\label{stationary}
\end{eqnarray}
It generates the stationary eigenvalue equation
\begin{eqnarray}
H\,\Phi = M\,\Phi\,,\qquad M = m_F+m_S+E
\end{eqnarray}

where the states of opposite parity have the form

\begin{eqnarray}
\Phi_{\mathrm{I\phantom{I}}}(\boldsymbol{r}) = 
\begin{pmatrix}
a_1(r)\,\,\Omega_{j\ell m}(\theta,\phi)\spazio{0.4}\\ 
a_2(r)\,\,\Omega_{j\ell' m}(\theta,\phi)\spazio{0.4}\\ 
a_3(r)\,\,\Omega_{j\ell m}(\theta,\phi),\spazio{0.4}\\
a_4(r)\,\,\Omega_{j\ell' m}(\theta,\phi)
\end{pmatrix}\,,
\qquad
\Phi_{\mathrm{II}}(\boldsymbol{r}) = 
\begin{pmatrix}
a_1(r)\,\,\Omega_{j\ell' m}(\theta,\phi)\spazio{0.4}\\ 
a_2(r)\,\,\Omega_{j\ell  m}(\theta,\phi)\spazio{0.4}\\ 
a_3(r)\,\,\Omega_{j\ell' m}(\theta,\phi)\spazio{0.4}\\
a_4(r)\,\,\Omega_{j\ell  m}(\theta,\phi)
\end{pmatrix}
\label{States}
\end{eqnarray}

and $a_i(r)$ are coefficients depending upon the relative radial variable only.
We redefine  two of the unknown functions by letting 
$\,\,a_2(r)\rightarrow -ia_2(r)\,,\quad$  $\,\,a_4(r)\rightarrow -ia_4(r)\,\,~$
and we introduce the dimensionless variables

\begin{eqnarray}
\boldsymbol{s}=m_F\,\boldsymbol{r},\quad~ \sigma=m_S/m_F,\quad~ \epsilon=E/m_F,\quad~ \lambda=M/m_F = 1+\sigma+\epsilon,\quad
\label{dimlessvar}
\end{eqnarray}

By substituting (\ref{States}) into (\ref{stationary}) 
we get a system of eight equations equal in pairs,
and the final system to be solved is
\begin{eqnarray}
&{}&\biggl(\,\frac{d}{ds}-\frac{j-1/2}{s}\,\biggr) \,a_1(s)
+\Bigl(\,\lambda+\frac{\alpha}{s}-\sigma +P \,\Bigr)\,a_2(s)-2\sigma\,a_4(s)=0\spazio{1.2}\cr 
&{}&\biggl(\,\frac{d}{ds}+\frac{j+3/2}{s}\,\biggr)\,a_2(s)
-\Bigl(\,\lambda+\frac{\alpha}{s}-\sigma - P \,\Bigr)\,a_1(s)+2\sigma\,a_3(s)=0\spazio{1.2}\cr
&{}&\biggl(\,\frac{d}{ds}+\frac{j+3/2}{s}\,\biggr) \,a_4(s)
-\Bigl(\,\lambda+\frac{\alpha}{s}+\sigma - P \,\Bigr)\,a_3(s)-
\frac{\nabla^2}{2\sigma}\,a_1(s)=0
\spazio{1.2}\cr
&{}&\biggl(\,\frac{d}{ds}-\frac{j-1/2}{s}\,\biggr) \,a_3(s)
+\Bigl(\,\lambda+\frac{\alpha}{s}+\sigma + P \,\Bigr)\,a_4(s)+
\frac{\nabla^2}{2\sigma}\,a_2(s)=0 
\label{SysEvenOdd}
\end{eqnarray}
where $\,P=\pm1\,$, the positive sign referring to the state $\,\Phi_{\mathrm{I}}(\boldsymbol{s})\,$, the negative sign to $\,\Phi_{\mathrm{II}}(\boldsymbol{s})\,$.
Notice that the Laplace operator of the third equation 
contains an angular momentum contribution 
$r^{-2}\,(j-1/2)(j+1/2)$, while in the last equation
the angular momentum contribution is $r^{-2}\,(j+1/2)(j+3/2)$.
Actually, from (\ref{SysEvenOdd}) it turns out that $a_3(r)$ and $a_4(r)$
can be expressed in terms of $a_1(r)$ and $a_2(r)$ by the 
algebraic relations, equal for both parities:
\begin{equation}
a_3(s)=\frac{\Bigl(\bigl((\lambda-\sigma)s+\alpha\bigr)^2-s^2\Bigr)\,a_1(s)-\alpha
	\, a_2(s) }{4s\,(\lambda s+\alpha)\,\sigma}, \qquad
a_4(s)=\frac{\alpha\, a_1(s) + \Bigl(\bigl((\lambda-\sigma)s+\alpha\bigr)^2-s^2\Bigr)\,a_2(s) }{4s\,(\lambda s+\alpha)\,\sigma}\,.
\end{equation}
Denoting for simplicity $f(s)=\,a_1(s)\,$, $g(s)=-\,a_2(s)\,$ 
and introducing the parameters
\begin{eqnarray}
\Lambda=\frac{\lambda}{2}+\frac{1-\sigma^2}{2\,\lambda},\qquad B=\frac{\alpha\,(1-\sigma^2)}{2\,\lambda^2},
\qquad \rho= \frac{\alpha}{\lambda}
\label{LamBrho}
\end{eqnarray}
the differential problem to be solved reduces to the 
second order boundary condition problem with eigenvalue $\epsilon$ 
(or $\lambda$)
\begin{eqnarray}
&{}&\frac{d}{ds}f(s)+\Bigl(\,-\frac{j}{s}+\frac{1}{2\,(s+\rho)}\,\Bigr)\,f(s) -
\Bigl(\,\Lambda+1+\frac{\alpha}{2\,s}-\frac{B}{s+\rho}\,\Bigr)
\,g(s)=0\spazio{1.0}\cr
&{}&\frac{d}{ds}g(s)+
\Bigl(\,\frac{j+1}{s}+\frac{1}{\,2\,(s+\rho)}\,\Bigr)\,g(s) + \Bigl(\,\Lambda-1+\frac{\alpha}{2\,s}-\frac{B}{s+\rho}\,\Bigr)\,f(s) 
=0
\label{Sys2dim}
\end{eqnarray}
The system (\ref{Sys2dim}) can be given a more
convenient form by substituting the angular 
momentum $\,j\,$ with $\,\kappa=-j-1/2\,$ as in
\cite{BLP} and by changing the unknown functions
$\,f(s),\,g(s)\,$ with $\,\phi(s),\,\chi(s)\,$ according to
\begin{eqnarray}
f(s)=\frac{\phi(s)}{\sqrt{s\,(s+\rho)}}, \qquad
g(s)=\frac{\chi(s)}{\sqrt{s\,(s+\rho)}}.
\label{phichi}
\end{eqnarray}
We get
\begin{eqnarray}
&{}&\frac{d}{ds}\phi(s)+\frac{\kappa}{s}\,\phi(s) -
\Bigl(\,\Lambda+1+\frac{\alpha}{2\,s}-\frac{B}{s+\rho}\,\Bigr)
\,\chi(s)=0\spazio{1.0}\cr
&{}&\frac{d}{ds}\chi(s)-
\frac{\kappa}{s}\,\,\chi(s) + \Bigl(\,\Lambda-1+\frac{\alpha}{2\,s}-\frac{B}{s+\rho}\,\Bigr)\,\phi(s) 
=0
\label{sysdirac}
\end{eqnarray}
\bigskip

The system (\ref{sysdirac}) is formally identical to a 
Dirac equation in a central potential \cite{BLP}. 
However, besides the fact that for a given value of the
parameter $\,\Lambda\,$ there are two determinations of the eigenvalue $\,\lambda\,$ we are concerned with, {\it{i.e.}} $\,\lambda = \Lambda\pm\bigl(\Lambda^2+\sigma^2-1)^{1/2}\,$,
the major point that forbids the system (\ref{sysdirac}) 
to be treated as a genuine Dirac
equation is the dependence of the parameter $\,B\,$ on $\,\lambda\,$ 
and hence on $\,\Lambda\,$. In particular, for (\ref{sysdirac})
the spectrum is no more dependent only upon the
absolute value of $\,\kappa\,$ as for the Dirac equation 
in a central potential \cite{BLP}. Thus
the level degeneracy is removed and the states
with $\,\pm\kappa\,$ appear separately,  although their difference 
is really small, as we can see from the numerical results. 
Finally notice that an actual Dirac equation in a Coulomb 
potential with coupling constant $\,\alpha/2\,$ is obtained 
for equal fermion and scalar masses: in this case 
$\,\sigma=1, ~ B=0\,$ and $\,\Lambda=\lambda/2,$.

For general masses, the solution of (\ref{sysdirac}) 
cannot be completely analytical. It is a boundary value 
problem with singularities at the origin and at infinity 
and we use a double shooting framework for obtaining the 
eigenvalues and the eigenfunctions. 
However, in order
to keep a high standard of accuracy in the results, we can push 
far enough the analytical development and determine the
exact recurrence relations for the approximate solutions 
needed at the two singular points. 
Since the system is regular in $(0,\infty)$, we calculate two 
more holomorphic series solutions, 
each of which centered sufficiently close to one of the 
singular points. We then match them to the approximate 
solution coming from the corresponding singular point. 
The continuity of components of these two series at an 
arbitrarily chosen crossing point gives, as usual, the 
spectral condition. 

We conclude this section with an observation on the limits 
of (\ref{SysEvenOdd}) for large values of the mass of each 
one of the two components.
Since $\lambda+\alpha/s-\sigma= 1+\epsilon+\alpha/s$,
when the mass of the scalar tends to infinity, namely 
$\sigma\rightarrow\infty$, from (\ref{dimlessvar})
we see that  the last two equations of (\ref{SysEvenOdd}) are 
identically vanishing and the two first ones
give the Dirac equation \cite{BLP}
\begin{eqnarray}
&{}&\Bigl(\,\frac{d}{ds}+\frac{1+\kappa}{s}\,\Bigr)\,a_1(s) +
\Bigl(\,1+\epsilon+\frac{\alpha}{s}+1\,\Bigr)\,a_2(s)=0\spazio{1.0}\cr
&{}&\Bigl(\,\frac{d}{ds}+\frac{1-\kappa}{s}\,\Bigr)\,a_2(s) -
\Bigl(\,1+\epsilon+\frac{\alpha}{s}-1\,\Bigr)\,a_1(s)=0
\label{Sys2limD}
\end{eqnarray}
When the mass of the fermion tends to infinity we recover 
the KG equation for the scalar.
This is easily seen by first making the change of variable 
$\,{\boldsymbol{s}} = (1/\sigma)\,{\boldsymbol{u}}\,$ and taking 
the mass of the lighter particle as mass scale. In order to 
normalize the physical parameters to the mass of the scalar,
 we define  $\,\eta = \epsilon/\sigma$.  Developing
$\,a_j(u)=(1/\sigma)\,a_j^{(1)}(u)+a_j^{(0)}(u),\,j=1..4\,$, 
at the order $\,\sigma^{-2}\,$ we find the vanishing of 
$\,a_2^{(1)}(u)\,$ and $\,a_4^{(1)}(u)\,$. At the order 
$\,\sigma^{-1}\,$ we then see that
$\,a_2^{(0)}(u),\,a_3^{(1)}(u)\,$ and $\,a_4^{(0)}(u)\,$ 
can be expressed in terms of $\,a_1^{(1)}(u)$,
which, in turn, satisfies 
\begin{equation}
\nabla^2\,a_1^{(1)}(u)+\biggl(\,\Bigl(1+\eta+\frac{\alpha}{u}\Bigr)^2-1\,\biggr)\,a_1^{(1)}(u) = 0
\end{equation}
namely the KG equation in a Coulomb field.

\medskip

%
\section{The solution of the boundary value problem.}
%
\medskip

In this section we give the details of our method for calculating 
the spectrum of the system (\ref{Sys2dim}), together with the results 
of the pure Coulomb levels of the
$\,({H}^+\!,\,\pi^-)\,$ and $\,({H}^+\!,\,{K}^-)\,$ atoms.
A global exact analytical solution is not available. Therefore 
we look for a result in the form of piecewise approximate analytical 
solutions expressed by series and, in in order to accelerate the 
convergence, by Pad\'e approximants.
This allows to obtain any required precision and to have 
an effective control of it.  
Clearly, any improvement of the accuracy occurs to the detriment 
of the computation time. 
However, numerical values of very large numbers of coefficients 
of the series involved are efficiently and rapidly found by 
recurrence procedures. Only the calculations of the Pad\'e are  
somewhat longer, still completely remaining within acceptable 
time limits even for a normal workstation.
Here below we give, thus, the explicit form of the 
recurrence relations of (\ref{sysdirac}) for the series
solution at the origin, the asymptotic solution at 
infinity and the holomorphic solution at any regular 
point of the open positive line $(0,\infty)$. 
\bigskip

\subsection*{\textbf{\textit{The solution at the origin.}}}\label{solzer}
\bigskip

For the solution at the origin we make the usual expansion of the unknown functions $\phi(s)$ and $\chi(s)$:
\begin{eqnarray}
\phi(s) = s^\nu\,\sum_{n=0}^{\infty}\phi_n\,s^n,\qquad 
\chi(s) = s^\nu\,\sum_{n=0}^{\infty}\chi_n\,s^n
\label{phichi0}
\end{eqnarray}

The index $\nu$ is obtained by the relations with $n=0$
\begin{eqnarray}
&{}&(\nu+\kappa)\,\phi_0-({\alpha}/2)\,\chi_0 =0\spazio{0.6}\cr
&{}&({\alpha}/2)\,\phi_0+(\nu-\kappa)\,\chi_0 =0
\end{eqnarray}
which actually are an eigenvalue system in $\nu$, yielding 
\begin{eqnarray}
\nu=\sqrt{\kappa^2-\alpha^2/4}, \qquad
\phi_0=1,\qquad \chi_0=(2/{\alpha})\,(\nu+\kappa)
\end{eqnarray}
By substituting (\ref{phichi0}) in the system (\ref{sysdirac}), 
expanded in the neighborhood of $s=0$, and letting
\begin{eqnarray}
\Sigma_{\phi,\,n}=\sum_{p=0}^{n}(-\rho)^{-p}\,\phi_{n-p},\qquad 
\Sigma_{\chi,\,n}=\sum_{p=0}^{n}(-\rho)^{-p}\,\chi_{n-p}
\end{eqnarray}
we obtain the  recurrence relations for $n\geq 1$. They read
\begin{eqnarray}
\begin{pmatrix}
\nu+\kappa+n & -\alpha/2 \spazio{0.8}\\
\alpha/2 & \nu-\kappa+n
\end{pmatrix}\,
\begin{pmatrix}
\phi_n\spazio{0.8}\\
\chi_n
\end{pmatrix} =
\begin{pmatrix}
\phantom{-}(\Lambda+1)\,\chi_{n-1}-(B/\rho)\,\Sigma_{\chi,n-1}
\spazio{0.8}\\
-(\Lambda-1)\,\phi_{n-1}+(B/\rho)\,\Sigma_{\phi,n-1}
\end{pmatrix}
\label{ricorre0}
\end{eqnarray}
Solving in $\phi_n$ and $\chi_n$, we get the expressions to 
be implemented by a numerical code:
\begin{eqnarray}
&{}& \phi_n={\displaystyle{\frac 1{2n\,(n+2\nu)}}}\,
\biggl(\,
\,2\,(\nu+n-\kappa)\,\Bigl(\,(\Lambda+1)\,\chi_{n-1}-
(B/\rho)\,\,\Sigma_{\chi,n-1}\,\Bigr) - \alpha\,\Bigl(\,(\Lambda-1)\,\phi_{n-1}-(B/\rho)\,\,\Sigma_{\phi,n-1}\,\Bigr)\,\biggr)\spazio{1.0}\cr
&{}& \chi_n={\displaystyle{\frac 1{2n\,(n+2\nu)}}}\,
\biggl(\,
-\alpha\,\Bigl(\,(\Lambda+1)\,\chi_{n-1}-(B/\rho)\,\,\Sigma_{\chi,n-1}\,\Bigr) - (\nu+\kappa+n)\,\Bigl(\,(\Lambda-1)\,\phi_{n-1}-(B/\rho)\,\,\Sigma_{\phi,n-1}\,\Bigr)\,\biggr)
\end{eqnarray}

The corresponding series have a finite radius of convergence $s_0$. 
A point in $(0,s_0)$ will be chosen to match this solution with the 
holomorphic series described here below.
\bigskip

\subsection*{\textbf{\textit{The solution at a regular point.}}}
\bigskip

We first shift the system (\ref{sysdirac}) at a point $\,u_0>0\,$ by 
letting $\,s=u+u_0\,$, so that $\,u=0\,$ is a regular point. We have
\begin{eqnarray}
&{}&\frac{d}{du}\phi(u)
+\frac{\kappa}{u+u_0}\,\phi(u) -
\Bigl(\,\Lambda+\frac{\alpha}{2\,(u+u_0)}
-\frac{B}{u+u_0+\rho}+1\,\Bigr)
\,\chi(u)=0\spazio{1.0}\cr
&{}&\frac{d}{du}\chi(u)-
\frac{\kappa}{u+u_0}\,\,\chi(u) + \Bigl(\,\Lambda+\frac{\alpha}{2\,(u+u_0)}
-\frac{B}{u+u_0+\rho}-1\,\Bigr)\,\phi(u) 
=0
\label{sysdiracu}
\end{eqnarray}
The expansion of the solutions at the regular point $\,u=0\,$ 
are now the holomorphic series
\begin{eqnarray}
\phi(u) = \sum_{n=0}^{\infty}\phi_n\,u^n,\qquad 
\chi(u) = \sum_{n=0}^{\infty}\chi_n\,u^n
\end{eqnarray}
The coefficients $\,(\phi_0,\chi_0)\,$ are arbitrary 
integration constants that will be chosen to be
the pairs $\,(1,0)\,$ and $\,(0,1)\,$  in order to
obtain a set of fundamental solutions, necessary for the matching 
with the expansions at the origin and at infinity. For $\,n\ge 1\,$, 
assuming that the coefficients with negative index are 
vanishing, a substitution into (\ref{sysdiracu}) yields the 
following expressions of $\phi_n$ and $\chi_n$
\begin{eqnarray}
&{}&\!\!\!\!\!\!\!\!\!\!\!\!\!\!\!\!
\phi_n =
{\displaystyle{\frac 1{n\,(u_0+\rho)\,u_0}}}\,
\biggl(\,
-\Bigl((\rho+2\,u_0)\,(n-1)+(\rho+u_0)\,\kappa\Bigr)\,\phi_{n-1}
+\Bigl((\Lambda+1)\,u_0^2+
\Bigl((\Lambda+1)\,\rho-B+\alpha/2\Bigr)\,u_0+\rho\,\alpha/2\Bigr)\,\chi_{n-1}
\cr
&{}&\phantom{\!\!\!\!\!\!\!\!\!\!\!\!\!\!\!\!\phi_n =
	{\displaystyle{\frac 1{n\,(u_0+\rho)\,u_0}}}\,
	\biggl(}
-\Bigl(\kappa+n-2\Bigr)\,\phi_{n-2}
+\Bigl((\rho+2\,u_0)\,(\Lambda+1)-B+\alpha/2\Bigr)\,\chi_{n-2}+(\Lambda+1)\,\chi_{n-3}\,
\biggr)\cr
&{}&\!\!\!\!\!\!\!\!\!\!\!\!\!\!\!\!
\chi_n =
{\displaystyle{\frac 1{n\,(u_0+\rho)\,u_0}}}\, \biggl(\,-\Bigl((\Lambda-1)\,u_0^2+\Bigl((\Lambda-1)\,\rho-B+\alpha/2\Bigr)\,u_0+\,\rho\,\alpha/2\Bigr)\,\phi_{n-1}
-\Bigl((\rho+2\,u_0)\,(n-1)-(\rho+u_0)\,\kappa\Bigr)\,\chi_{n-1}\cr
&{}&\phantom{\!\!\!\!\!\!\!\!\!\!\!\!\!\!\!\!\phi_n =
	{\displaystyle{\frac 1{n\,(u_0+\rho)\,u_0}}}\,
	\biggl(}
-\Bigl((\Lambda-1)\,(\rho+2\,u_0)-B+\alpha/2\Bigr)\,\phi_{n-2}+\Bigl(\kappa-n+2\Bigr) \,\chi_{n-2}-(\Lambda-1)\,\phi_{n-3}
\,\biggr)\,,
\end{eqnarray}
by which accurate numerical values of the solutions are rapidly obtained.
\bigskip

\subsection*{\textbf{\textit{The solution at infinity.}}}\label{solinf}
\bigskip

General theorems \cite{Cope} state that asymptotic solutions at infinity of (\ref{sysdirac}) are of the form
\begin{eqnarray}
\phi(s) = e^{-\omega s}\,
s^\delta\,\Phi(s), \qquad
\chi(s) = e^{-\omega s}\,
s^\delta\,X(s)
\label{asysol}
\end{eqnarray}
The unknown functions $\,(\Phi(s),X(s))\,$ are solutions of the system
\begin{eqnarray}
&{}&\frac{d}{ds}\Phi(s)+\Bigl(-\omega+\frac{\delta+\kappa}{s}\Bigr)\,\Phi(s) -
\Bigl(\,\Lambda+\frac{\alpha}{2\,s}-\frac{B}{s+\rho}+1\,\Bigr)
\,X(s)=0\spazio{1.0}\cr
&{}&\frac{d}{ds}X(s)+\Bigl(-\omega+\frac{\delta-\kappa}{s}\Bigr)\,X(s) + \Bigl(\,\Lambda+\frac{\alpha}{2\,s}-\frac{B}{s+\rho}-1\,\Bigr)\,\Phi(s) 
=0
\label{sysdiracsai}
\end{eqnarray}
By substituting into (\ref{sysdiracsai}) the expansions
\begin{eqnarray}
\Phi(s) = \sum_{n=0}^{\infty}\Phi_n\,s^{-n},\qquad 
X(s) = \sum_{n=0}^{\infty}X_n\,s^{-n}\qquad
(s+\rho)^{-1} =\sum_{n=0}^{\infty}(-\rho)^n\,s^{-(n+1)}
\label{seriegeom}
\end{eqnarray}
at the zero order we find the system
\begin{eqnarray}
\begin{pmatrix}\,
-\omega & -\Lambda-1\, \spazio{0.4}\\
\,\Lambda-1 & -\omega\,
\end{pmatrix}\,
\begin{pmatrix}\, \Phi_0\spazio{0.4}\\X_0\,\end{pmatrix}=0
\label{asiserie0}
\end{eqnarray}
which yields
\begin{eqnarray}
\omega = \sqrt{1-\Lambda^2}, \qquad \Phi_0 = 1, \qquad X_0 = -\omega/(1+\Lambda)
\label{muasi}
\end{eqnarray}
Letting
\begin{eqnarray}
\Sigma_{\Phi,\,n}=\sum_{p=0}^{n}(-\rho)^{p}\,\Phi_{n-p},\qquad 
\Sigma_{X,\,n}=\sum_{p=0}^{n}(-\rho)^{p}\,X_{n-p}
\end{eqnarray}
at order $\,-n\,$, with $\,n\geq 1\,$  we find the relations

\begin{eqnarray}
\begin{pmatrix}\,
-\omega & -\Lambda-1\, \spazio{0.4}\\
\,\Lambda-1 & -\omega\,
\end{pmatrix}\,
\begin{pmatrix}\, \Phi_n\spazio{0.8}\\X_n\,\end{pmatrix}=
\begin{pmatrix}\,
-(\delta+\kappa-n+1)\,\Phi_{n-1}+(\alpha/2)\,X_{n-1}-B
\,\Sigma_{X,\,n-1}
\spazio{0.8}\\
-(\delta-\kappa-n+1)\,X_{n-1}-(\alpha/2)\,\Phi_{n-1}+B
\,\Sigma_{\Phi,\,n-1}
\end{pmatrix}\,
\label{asiseriemn}
\end{eqnarray}
\medskip

The matrix in (\ref{asiseriemn}) is singular in view 
of (\ref{asiserie0}), with eigenvalues $\,(\,0,\,-2\omega\,)\,$. 
Therefore we transform the system to the basis where the matrix 
is diagonal. Obviously, in such a basis,
one of the two relations obtained from (\ref{asiseriemn}) is homogeneous. 
For $\,n=1\,$ this homogeneous equation gives the value of $\,\delta\,$, namely
\begin{eqnarray}
\delta = -(2\,\omega)^{-1}\,\Bigl(\Lambda\,(2B-\alpha)\Bigr)
\end{eqnarray}

In order to obtain a solvable set of recurrence
equations allowing to determine the coefficients
$\,\Phi_n,\,X_n\,$ for $\,n\geq 1\,$, we have 
considered a first equation by 
subtracting the homogeneous relation from the
inhomogeneous one, and a second equation obtained
by  rescaling the homogeneous
equation from $\,n\,$ to $\,n+1\,$, which thus becomes an inhomogeneous equation in $\,\Phi_n\,$ and $\,X_n\,$. The result is
\begin{eqnarray}
&{}&\Phi_n= \Delta^{-1}\,\biggl(\,
-A_X\,C_\Phi~\Phi_{n-1}-A_X\,C_X~X_{n-1}+\Upsilon\,\Bigl(2\,\Upsilon\,\omega\,B\,\rho-A_X\,B\,\Bigr)~\Sigma_{\Phi,n-1}+\Bigl(2\,\Upsilon\,\omega\,B\,\rho+A_X\,B\Bigr)~\Sigma_{X,n-1}\,\biggr)\spazio{1.0}\cr
&{}&X_n= \Delta^{-1}\,\biggl(\,\phantom{-}
A_\Phi\,C_\Phi~\Phi_{n-1}+A_\Phi\,C_X~X_{n-1}+\Upsilon\,\Bigl(-2\,\omega\,B\,\rho+A_\Phi\,B\,\Bigr)~\Sigma_{\Phi,n-1}-(2\,\omega\,B\,\rho+A_\Phi\,B\Bigr)~\Sigma_{X,n-1}
\,\biggr)
\end{eqnarray}
where, for simplicity in writing, we have defined
\begin{eqnarray} 
&{}&A_\Phi = n-\delta-\kappa-(2\,B-\alpha)\,(\Lambda+1)/(2\,\omega), \qquad A_X = (1/2)\,\alpha-(\Lambda+1)\,(n-\delta+\kappa)/\omega-B, \spazio{0.8}\cr
&{}&\phantom{2\omega)}C_\Phi = n-1-\kappa-\delta-(\Lambda+1)\,\alpha/(2\,\omega), \qquad C_X = (1/2)\,\alpha+(\Lambda+1)\,(n-1-\delta+\kappa)/\omega, \spazio{0.8}\cr
&{}&\phantom{XXXXXXXXXXXX}\phantom{(2\omega),}
\Upsilon = (\Lambda+1)/\omega, \qquad 
\phantom{{}_X}\Delta = (-2\,A_\Phi\,\Upsilon+2\,A_X)\,\omega
\end{eqnarray}
It is well known that the expansion of the solution at 
infinity produces an asymptotic series. Thus, for convergence, 
in the numerical treatment we have applied
to this series the Pad\'e approximant technique.

The results for the pure Coulomb levels of the pionic and 
kaonic atoms are reported in the following Tables 1 and 2.
\bigskip\medskip

%
%
%
\begin{table}[h]
	\begin{center}
		\begin{small}
			{{ \begin{tabular}{|c|c||r|r|r|r|r|}
						\hline 
						State
						&$\kappa$
						&$n=1$\phantom{9011}
						&$n=2$\phantom{901.}
						&$n=3$\phantom{901.}
						&$n=4$\phantom{901.}
						&$n=5$\phantom{901.}
						\\
						\hline 
						$s_{1/2}$
						&$-1$ 
						&-3235.0931859
						&-808.7611897
						&-359.4465684
						&-202.1877820
						&-129.3998072
						\\
						\hline
						$p_{1/2}$
						&$+1$ 
						&-\phantom{09319}
						&-808.7426715
						&-359.4410815
						&-202.1854673
						&-129.3986221
						\\ 
						$p_{3/2}$
						&$-2$ 
						&-\phantom{09319}
						&-808.7424909
						&-359.4410280
						&-202.1854447
						&-129.3986105
						\\ 
						\hline		 
						$d_{3/2}$
						&$+2$
						&-\phantom{09319}
						&-\phantom{09319} 					  
						&-359.4399305
						&-202.1849817
						&-129.3983734
						\\
						$d_{5/2}$
						&$-3$
						&-\phantom{09319}
						&-\phantom{09319} 					  
						&-359.4399127 
						&-202.1849742 
						&-129.3983696
						\\
						\hline 
						$f_{5/2}$
						&$+3$ 
						&-\phantom{09319}
						&-\phantom{09319}
						&-\phantom{09319}
						&-202.1847757
						&-129.3982680
						\\ 
						$f_{7/2}$
						&$-4$ 
						&-\phantom{09319}
						&-\phantom{09319} 
						&-\phantom{09319}
						&-202.1847720
						&-129.3982661
						\\
						\hline 
						$g_{7/2}$
						&$+4$ 
						&-\phantom{09319}
						&-\phantom{09319} 
						&-\phantom{09319}
						&-\phantom{09319}
						&-129.3982096
						\\ 
						$g_{9/2}$ 
						&$-5$
						&-\phantom{09319}
						&-\phantom{09319} 
						&-\phantom{09319}
						&-\phantom{09319}
						&-129.3982085
						\\ 
						\hline
					\end{tabular}
			}}
		\end{small}
		\caption{The pure Coulomb spectrum of the pionic atom $\,({H}^+\!,\,\pi^-)\,$ expressed in eV. We
			have taken the proton mass $m_P$=938.2720813 MeV, the pion mass $m_\pi$=139.57061 MeV,
			the fine structure constant $\alpha$=0.0072973525698163 \cite{PDG}.
		  In the first row $n=n_r+|\kappa|$ is the
	  principal quantum number, where $n_r$ is the
  radial quantum number  \cite{BLP}.}
	\end{center}
\end{table}
%
%
%
%
%
%
\begin{table}[h]
	\begin{center}
		\begin{small}
			{{ \begin{tabular}{|c|c||r|r|r|r|r|}
						\hline 
						State
						&$\kappa$
						&$n=1$\phantom{9011}
						&$n=2$\phantom{901.}
						&$n=3$\phantom{901.}
						&$n=4$\phantom{901.}
						&$n=5$\phantom{901.}
						\\
						\hline 
						$s_{1/2}$
						&$-1$ 
						&-8612.9384350
						&-2153.2252598
						&-956.9862506
						&-538.3038471
						&-344.5140798
						\\
						\hline
						$p_{1/2}$
						&$+1$ 
						&-\phantom{09319}
						&-2153.2097136
						&-956.9816444
						&-538.3019039
						&-344.5130848
						\\ 
						$p_{3/2}$
						&$-2$ 
						&-\phantom{09319}
						&-2153.2063064
						&-956.9806348
						&-538.3014780
						&-344.5128669
						\\ 
						\hline		 
						$d_{3/2}$
						&$+2$
						&-\phantom{09319}
						&-\phantom{09319} 					  
						&-956.9797133
						&-538.3010892
						&-344.5126677
						\\
						$d_{5/2}$
						&$-3$
						&-\phantom{09319}
						&-\phantom{09319} 					  
						&-956.9793768
						&-538.3009472
						&-344.5125950
						\\
						\hline 
						$f_{5/2}$
						&$+3$ 
						&-\phantom{09319}
						&-\phantom{09319}
						&-\phantom{09319}
						&-538.3007806
						&-344.5125097
						\\ 
						$f_{7/2}$
						&$-4$ 
						&-\phantom{09319}
						&-\phantom{09319} 
						&-\phantom{09319}
						&-538.3007096
						&-344.5124734
						\\
						\hline 
						$g_{7/2}$
						&$+4$ 
						&-\phantom{09319}
						&-\phantom{09319} 
						&-\phantom{09319}
						&-\phantom{09319}
						&-344.5124260
						\\ 
						$g_{9/2}$ 
						&$-5$
						&-\phantom{09319}
						&-\phantom{09319} 
						&-\phantom{09319}
						&-\phantom{09319}
						&-344.5124042
						\\ 
						\hline
					\end{tabular}
			}}
		\end{small}
		\caption{The pure Coulomb spectrum of the kaonic atom $\,({H}^+\!,\,K^-)\,$ expressed in eV. 
			$m_P$ and $\alpha$ are as in Table 1. We have taken the mass of the $K$ meson $m_K$=493.677 MeV \cite{PDG}.
		In the first row $n=n_r+|\kappa|$ is the
		principal quantum number, where $n_r$ is the
		radial quantum number  \cite{BLP}.}
	\end{center}
\end{table}
%

%
\section{The lowest order radiative corrections}
%
\medskip

The lowest order QED corrections to the spectrum is given 
by the one-loop electron vacuum polarization or 
Uehling potential. It takes into account the vacuum polarization 
produced by the creation of electron-positron pairs,
since the contributions due to other particles, such as 
the lepton $\,\mu\,$, are much smaller. 
Its form, in terms of the dimensionless variable $s$ and 
expressed in proton mass units, is
\begin{equation}
V_U(s) = -\frac{2\alpha^2}{3\pi s}\,\int\limits_1^\infty {\rm{e}}^{\displaystyle{-\Bigl(\,\frac{2m_e}{m_P}\,s x\,\Bigr)}}\,\sqrt{x^2-1}~\frac{2x^2+1}{2x^4}~dx
\end{equation}
where $m_P$ is the proton mass and $m_e$ the electron mass. 
It can be expressed in terms of special functions
\begin{equation}
V_U(s) = -\frac{2\alpha^2}{3\pi s}\,\left(\,\frac{m_e s}{4m_P}\,G^{30}_{13}\,\biggl({{1}\atop{0,-1/2,-1/2}}\,
\biggl |\,
\Bigl(\frac{m_e}{m_P}\,s\Bigr)^2\biggr)+ \Bigl(\frac{m_e s}{4m_P}\Bigr)^3 \,G^{30}_{13}\,\biggl({{1}\atop{-1/2,-1,-3/2}}\biggr)
\,
\biggl |\,
\Bigl(\frac{m_e}{m_P}\,s\Bigr)^2\,\right)
\end{equation}
where
$\,G^{mn}_{pq}\bigl({{a_1,...,a_p}\atop{b_1,...,b_q}}\,|\,z\bigr)$
is the Meijer G function \cite{BS}.

The next order corrections are given by the 
two-loop irreducible part of the the vacuum polarization, 
also known as K\"allen-Sabry potential
\begin{eqnarray}
&{}&\!\!\!\!\!\!\!\!\!\!\!\!
V_{KS}(s)=-\frac{2\alpha^3}{3\pi^2s}\int\limits_{0}^{1}
\frac{{x\,\rm{e}}^{\displaystyle{-\Bigl(\,\frac{2m_e}{m_P}\frac s{\sqrt{1-x^2}}\,\Bigr)}}}{(1-x^2)}\,
\biggl[ x\,(3-x^2)\,\biggl(\frac32\ln\Bigl(\frac{1-x^2}{4}\Bigr)-2\,\ln(x)\biggr)+
\ln\Bigl(\frac{1+x}{1-x}\Bigr)\,\Bigl(\frac{11}{16}\,(3-x^2)\,(1+x^2)+\frac{x^4}{4}\Bigr)+
\spazio{1.0}\cr
&{}&\phantom{xxxxx}
(3-x^2)(1+x^2)\,\biggl({\rm{Li_2}}\Bigl(-\frac{1-x}{1+x}\Bigr)+2{\rm{Li_2}}\Bigl(\frac{1-x}{1+x}\Bigr)+
\ln\Bigl(\frac{1+x}{1-x}\Bigr)\,\Bigl(\frac32\ln\Bigl(\frac{1+x}{2}\Bigr)-\ln(x)\Bigr)\biggr)+\frac38 x\,(5-3\,x^2)\biggr]\,dx
\label{KSpot}
\end{eqnarray}
where $\,{\rm{Li_2}}(z)\,$ is the Dilogarithm function. 

We finally add the contribution of the reducible part of 
the two-loop vacuum polarization potential, namely
\begin{equation}
V_{R}(s) = -\frac{\alpha^3}{9\pi^2s}\int\limits_{0}^{1}
\frac{{x^2\,\rm{e}}^{\displaystyle{-\Bigl(\,\frac{2m_e}{m_P}\frac s{\sqrt{1-x^2}}\,\Bigr)}}}{(1-x^2)}\,
\Bigl(1-\frac{x^2}{3}\Bigl)\,\biggl(16-6x^2+3x\,(3-x^2)\,\ln\Bigl(\frac{1-x}{1+x}\Bigr)\biggr)
\label{Rpot}
%
\end{equation}
The detailed results of each one of the corrections on each 
atomic level are shown in Appendix II, together with some
comments. Here below we give the QED-corrected
transitions from the excited to the ground state
for $(H^+,\pi^-)$ and $(H^+,K^-)$ 
\bigskip

%
\begin{table}[h]
	\begin{center}
		\begin{small}
			{{ \begin{tabular}{|c||r|r|r|r|}
						\hline 
						Transition
						&$n=2\phantom{110}$ 
						&$n=3\phantom{110}$
						&$n=4\phantom{110}$
						&$n=5\phantom{110}$
						\\
						\hline					
						${}^ns_{1/2}\rightarrow {}^1s_{1/2}$
						&2429.2202909
						&2878.7969532
						&3036.1184511
						&3108.9286213
						\\
						\hline
						${}^np_{3/2}\rightarrow {}^1s_{1/2}$
						&2429.5729174
						&2878.8990487
						&3036.1611897
						&3108.9504262
						\\
						\hline 
						${}^nd_{5/2}\rightarrow {}^1s_{1/2}$
						&-\phantom{09319}
						&2878.9112441
						&3036.1663831
						&3108.9530963
						\\ 
						\hline		 
						${}^nf_{7/2}\rightarrow {}^1s_{1/2}$
						&-\phantom{09319}
						&-\phantom{09319}
						&3036.1668317
						&3108.9533385
						\\ 
						\hline
						${}^ng_{9/2}\rightarrow {}^1s_{1/2}$
						&-\phantom{09319}
						&-\phantom{09319}
						&-\phantom{09319}
						&3108.9533999
						\\ 
						\hline
					\end{tabular}
			}}
		\end{small}
		\caption{The values in eV of the transitions from the excited to the ground
			state of $(H^+,\pi^-)$. 
		The QED corrected value of the ground level
	is $\epsilon_{0,\,(H^+,\pi^-)}=-3238.3591996$ eV. }
	\end{center}
\end{table}
%
%

%
%
%
\begin{table}[h]
	\begin{center}
		\begin{small}
			{{ \begin{tabular}{|c||r|r|r|r|r|}
						\hline 
						Transition
						&$n=2\phantom{110}$ 
						&$n=3\phantom{110}$
						&$n=4\phantom{110}$
						&$n=5\phantom{110}$
						\\
						\hline					
						${}^ns_{1/2}\rightarrow {}^1s_{1/2}$
						&6479.1535386
						&7677.1203701
						&8096.2083183
						&8290.1408666
						\\
						\hline
						${}^np_{3/2}\rightarrow {}^1s_{1/2}$
						&6480.8162040
						&7677.6072875
						&8096.4123161
						&8290.2449542
						\\
						\hline 
						${}^nd_{5/2}\rightarrow {}^1s_{1/2}$
						&-\phantom{09319}
						&7677.7851662
						&8096.4842761
						&8290.2811709
						\\ 
						\hline		 
						${}^nf_{7/2}\rightarrow {}^1s_{1/2}$
						&-\phantom{09319}
						&-\phantom{09319}
						&8096.5009525
						&8290.2897706
						\\ 
						\hline
						${}^ng_{9/2}\rightarrow {}^1s_{1/2}$
						&-\phantom{09319}
						&-\phantom{09319}
						&-\phantom{09319}
						&8290.2910456
						\\ 
						\hline
					\end{tabular}
			}}
		\end{small}
		\caption{The values in eV of the transitions from the excited to the ground
			state of $(H^+,K^-)$.
			The QED corrected value of the ground level
		is $\epsilon_{0,\,(H^+,K^-)}=-8634.8934971$ eV. }
	\end{center}
\end{table}
We have done a comparison with experimental evaluations 
of the transitions between higher levels
of the pionic atom and we have found complete agreement
 within the experimental errors 
\cite{Hetal,Getal}. The experimental results for
kaonic atom are more rare and affected by larger errors \cite{cata}. 
There is however agreement between our
numerical values and those reported in \cite{cata}
taken from \cite{iwa}.

%
\section{Conclusions.}
%
\bigskip

In this paper we have calculated levels and states,
up to $n=5$,  of the pionic and kaonic Hydrogen
atoms in a two body quantum relativistic framework
for a scalar and a fermion. 
The original four dimensional system is reduced to a
pair of first order differential equations formally
identical to a Dirac equation in a central
potential, but for the eigenvalue. This fact, together with
the not purely Coulomb interaction,  implies  the 
non conservation of the Johnson-Lippmann operator
\cite{JL} and thus the splitting of the levels
having the same $n$ and $\kappa$ of equal absolute value 
but opposite sign. This is quite evident for 
$n=2$,  much less for increasing values of $n$ 
and $|\kappa|$,  as shown in Tables 1
and 2. The radiative corrections, Table 5, share 
the same behavior, with the hints concerning the
contributions  due to $V_R$.
The numerical calculations have been carried on 
with high accuracy, in order to give a clearcut
distinction and  enumeration of the levels although,
nowadays, they still cannot be experimentally observed. 
In any case the  great physical relevance of the 
mesic atoms is well known: for instance
many decades ago Dalitz  underlined the great
importance of the ``definitive determination of the 
energy level shifts in the $K$-Hydrogen and Deuterium, 
because of their direct connection with the physics of 
$K$-nucleon interaction and their complete independence 
from all other kinds of measurements which bear on this 
interaction'' (quoted by \cite{Curc}). 
This justifies the efforts in order to obtain
the electromagnetic spectrum as precise as possible.
The experiments now in progress  mainly in kaonic atoms 
\cite{Curc,Marton} will hopefully give a breakthrough in 
this direction.

\bigskip

%
%
\section*{Appendix I. $~$Variables, kinematics, dynamics.}
%
\bigskip

In previous papers we have explained in detail
the kinematic setting and the covariance of the treatment. 
Referring to those papers for details, we report here, for
the sake of completeness, a brief synthesis useful to 
contextualize the definitions of dimensional and dimensionless 
variables which appear in our equations.
We are given the phase space of two relativistic points, 
spanned by the Minkowski coordinates
$\,x_1^\mu,\,x_2^\mu$ and conjugate momenta $\,p_1^\mu,\,p_2^\mu$, 
the Lorentz metric tensor being $\eta^\mu_\nu={{\rm diag}}(1,-1,-1,-1).$ 
It is first instrumental to introduce a canonical transformation to 
`relative' and `global' coordinates
$\,\tilde{r}^\mu=x_1^\mu-x_2^\mu,\,\,X^\mu=(1/2)\,(x_1^\mu+x_2^\mu),$ 
with conjugate momenta
$\,\tilde{q}^\mu=(1/2)\,(p_1^\mu-p_2^\mu),\,\,P^\mu=p_1^\mu+p_2^\mu$.
We then make a second canonical transformation defined by 
its action on the momenta: it acts as the identity on $P^\mu$ 
and transforms  $\tilde{q}^\mu$ by boosting it
to the $\boldsymbol{P}=0$ frame, thus defining 
$\,q_\alpha=\varepsilon_\alpha^\mu(P)\,\tilde{q}_\alpha$, where
$\,\alpha=(0,A),$  $A=1,2,3\,$ and 
$\,\varepsilon_\alpha^\mu(P)=L^{-1}(\boldsymbol{P})_\alpha^\mu$ is
the matrix of the Lorentz boost.
This is a point-like transformation easily completed 
by the generating function 
yielding the corresponding coordinates conjugate to 
$q_\alpha$ and $P^\mu$, namely: $\,r_\alpha=\varepsilon_\alpha^\mu(P)\,\tilde{r}_\alpha\,$ and 
$\,Z^\mu=X^\mu+
P^{-2}\,\bigl(P^\nu\tilde{q}_\nu\,\tilde{r}^\mu-P^\nu\tilde{r}_\nu\,\tilde{q}^\mu+(P_0+P)^{-1}\,W^{0\mu}\bigr),\,$ where $\,W^{\mu\nu}\,$ 
is the Pauli-Lubanski tensor.
It is then obvious that $\,q_0\,$ and  $\,r_0\,$ are Lorentz invariant such as
$\,q=\bigl(q_A\,q_A\bigr)^{1/2},~ r=\bigl(r_A\,r_A\bigr)^{1/2},\,$ since
$\,\boldsymbol{q}=\{q_A\}\,$ and  $\,\boldsymbol{r}=\{r_A\}\,$ 
are Wigner vectors. Moreover $\,\emph{\textbf{Z}}\,$ is a position 
vector of the Newton-Wigner type and $\,Z^0\,$ has the covariance 
of a Lorentz time. The essential point is that the mass
shell conditions $\,p_1^2= m_1^2\,$ and $\,p_2^2= m_2^2\,$ expressed in these
coordinates, amount to the elimination of $\,q_0\,$ from the resulting energy 
(and Hamiltonian) of the relative system: $\,(m_1^2+ \boldsymbol{q}^2)^{1/2} 
+ (m_2^2+ \boldsymbol{q}^2)^{1/2},\,$ which plays the role 
of the invariant mass.
We then see that $\,Z^0\,$ remains the only one time in the system, since the
relative time $\,r_0\,$ is a cyclic variable and disappears: we have therefore
realized a canonical reduction of the phase space where 
$\,m_1\,$ and $\,m_2\,$ appear as fixed parameters, overcoming 
the problem of the relative energy which 
becomes an irrelevant quantity. The cyclic time  $\,r_0\,$  assumes the role 
of a gauge-type variable, arbitrarily  fixed in order to
recover separate descriptions of the two particles world lines.
The physical mass spectrum is determined by the 
invariant Hamiltonian completed with  invariant interaction terms depending on 
$\,\boldsymbol{q}\,$  and $\,\boldsymbol{r}.$ 
Thus the action of the Poincar\'e group is well 
defined and the dynamics is covariant.
Finally, the quantization is done in the tensor product 
of the particle Hilbert spaces,
using gamma matrices for fermions and an adapted Feshbach-Villars 
representation for the scalars, so that the square roots are 
eliminated from $\,H\,$ in any case.

\vfil\break

\section*{Appendix II. $~$The lowest order radiative correction.}
\bigskip

\renewcommand{\theequation}{II.\arabic{equation}}
We report here the detailed list of the QED corrections 
to the levels of Tables 1 and 2.

\begin{table}[h]
	\begin{center}
		\begin{small}$\quad$
			\!\!\!\!\!\begin{tabular}{|c|c||r|r|r||r|r|r|}
				\hline 
				State
				\!\!&\!\!$\kappa$
				\!\!&\!\!$V_U~{(H^+,\,\pi^-)}$\phantom{90}
				\!\!&\!\!$V_{KS}~{(H^+,\,\pi^-)}$\phantom{90}
				\!\!&\!\!$V_{R}~{(H^+,\,\pi^-)}$\phantom{90}
				\!\!&\!\!$V_{U}~{(H^+,\,K^-)}$\phantom{90}
				\!\!&\!\!$V_{KS}~{(H^+,\,K^-)}$\phantom{90}
				\!\!&\!\!$V_{R}~{(H^+,\,K^-)}$\phantom{90}
				\\
				\hline 
				$1s_{1/2}$ 
				\!\!&\!\!$-1$
				\!\!&\!\!-3.2413174\phantom{$\cdot{10^{-1}}$}
				\!\!&\!\!-2.0900666$\cdot{10^{-2}}$
				\!\!&\!\! 3.7955745$\cdot{10^{-3}}$
				\!\!&\!\!-21.7953398\phantom{$\cdot{10^{-3}}$}
				\!\!&\!\!-1.1473759$\cdot{10^{-1}}$
				\!\!&\!\! 4.4984666$\cdot{10^{-2}}$
				\\
				\hline
				$2s_{1/2}$
				\!\!&\!\!$-1$ 
				\!\!&\!\!-3.6831740$\cdot{10^{-1}}$
				\!\!&\!\!-2.2911091$\cdot{10^{-3}}$ 
				\!\!&\!\! 4.8071080$\cdot{10^{-4}}$   
				\!\!&\!\!-2.4169328\phantom{$\cdot{10^{-3}}$}
				\!\!&\!\!-1.2919109$\cdot{10^{-2}}$
				\!\!&\!\! 5.1225963$\cdot{10^{-3}}$      
				\\
				%
				$2p_{1/2}$
				\!\!&\!\!$+1$ 
				\!\!&\!\!-3.5796665$\cdot{10^{-2}}$
				\!\!&\!\!-3.7779853$\cdot{10^{-4}}$ 
				\!\!&\!\!-2.5718595$\cdot{10^{-5}}$          
				\!\!&\!\!-7.7448229$\cdot{10^{-1}}$
				\!\!&\!\!-6.5377131$\cdot{10^{-3}}$
				\!\!&\!\!-8.4622213$\cdot{10^{-6}}$      
				\\
				%
				$2p_{3/2}$
				\!\!&\!\!$-2$ 
				\!\!&\!\!-3.5796506$\cdot{10^{-2}}$
				\!\!&\!\!-3.7779729$\cdot{10^{-4}}$
				\!\!&\!\!-2.5718666$\cdot{10^{-5}}$           
				\!\!&\!\!-7.7447115$\cdot{10^{-1}}$
				\!\!&\!\!-6.5376417$\cdot{10^{-3}}$
				\!\!&\!\!-8.4748167$\cdot{10^{-6}}$      
				\\
				\hline
				$3s_{1/2}$
				\!\!&\!\!$-1$ 
				\!\!&\!\!-1.0756349$\cdot{10^{-1}}$
				\!\!&\!\!-6.6578532$\cdot{10^{-4}}$
				\!\!&\!\! 1.4242775$\cdot{10^{-4}}$          
				\!\!&\!\!-6.9476605$\cdot{10^{-1}}$
				\!\!&\!\!-3.6514009$\cdot{10^{-3}}$
				\!\!&\!\! 1.5104177$\cdot{10^{-3}}$     
				\\ 
				%
				$3p_{1/2}$
				\!\!&\!\!$+1$ 
				\!\!&\!\!-1.1407162$\cdot{10^{-2}}$
				\!\!&\!\!-1.1722417$\cdot{10^{-4}}$
				\!\!&\!\!-7.3772433$\cdot{10^{-6}}$           
				\!\!&\!\!-2.1392561$\cdot{10^{-1}}$
				\!\!&\!\!-1.7166683$\cdot{10^{-3}}$
				\!\!&\!\! 3.3703260$\cdot{10^{-5}}$      
				\\
				%
				$3p_{3/2}$
				\!\!&\!\!$-2$ 
				\!\!&\!\!-1.1407110$\cdot{10^{-2}}$
				\!\!&\!\!-1.1722378$\cdot{10^{-4}}$
				\!\!&\!\!-7.3772720$\cdot{10^{-6}}$           
				\!\!&\!\!-2.1392247$\cdot{10^{-1}}$
				\!\!&\!\!-1.7166496$\cdot{10^{-3}}$
				\!\!&\!\! 3.3698833$\cdot{10^{-5}}$      
				\\
				%
				$3d_{3/2}$
				\!\!&\!\!$+2$ 
				\!\!&\!\!-4.4463586$\cdot{10^{-4}}$
				\!\!&\!\!-6.3140024$\cdot{10^{-6}}$
				\!\!&\!\!-7.0773072$\cdot{10^{-7}}$          
				\!\!&\!\!-3.8501657$\cdot{10^{-2}}$
				\!\!&\!\!-4.4371663$\cdot{10^{-4}}$
				\!\!&\!\!-3.9638974$\cdot{10^{-5}}$     
				\\ 
				%
				$3d_{5/2}$
				\!\!&\!\!$-3$ 
				\!\!&\!\!-4.4463503$\cdot{10^{-4}}$
				\!\!&\!\!-6.3139928$\cdot{10^{-6}}$
				\!\!&\!\!-7.0772989$\cdot{10^{-7}}$           
				\!\!&\!\!-3.8501463$\cdot{10^{-2}}$
				\!\!&\!\!-4.4371479$\cdot{10^{-4}}$
				\!\!&\!\!-3.9638894$\cdot{10^{-5}}$      
				\\
				\hline
				$4s_{1/2}$
				\!\!&\!\!$-1$ 
				\!\!&\!\!-4.5156315$\cdot{10^{-2}}$
				\!\!&\!\!-2.7904635$\cdot{10^{-4}}$
				\!\!&\!\! 6.0079123$\cdot{10^{-5}}$           
				\!\!&\!\!-2.9047738$\cdot{10^{-1}}$
				\!\!&\!\!-1.5204611$\cdot{10^{-3}}$
				\!\!&\!\!6.3556694$\cdot{10^{-4}}$     
				\\
				%
				$4p_{1/2}$
				\!\!&\!\!$+1$ 
				\!\!&\!\!-4.9207863$\cdot{10^{-3}}$
				\!\!&\!\!-5.0167795$\cdot{10^{-5}}$
				\!\!&\!\!-3.0743111$\cdot{10^{-6}}$           
				\!\!&\!\!-8.9047505$\cdot{10^{-2}}$
				\!\!&\!\!-7.0538006$\cdot{10^{-4}}$
				\!\!&\!\! 1.7877271$\cdot{10^{-5}}$      
				\\
				%
				$4p_{3/2}$
				\!\!&\!\!$-2$ 
				\!\!&\!\!-4.9207642$\cdot{10^{-3}}$
				\!\!&\!\!-5.0167631$\cdot{10^{-5}}$
				\!\!&\!\!-3.0743244$\cdot{10^{-6}}$            
				\!\!&\!\!-8.9046201$\cdot{10^{-2}}$
				\!\!&\!\!-7.0537249$\cdot{10^{-4}}$
				\!\!&\!\! 1.7875322$\cdot{10^{-5}}$       
				\\
				%
				$4d_{3/2}$
				\!\!&\!\!$+2$ 
				\!\!&\!\!-2.4729903$\cdot{10^{-4}}$
				\!\!&\!\!-3.4598119$\cdot{10^{-6}}$
				\!\!&\!\!-3.8498620$\cdot{10^{-7}}$            
				\!\!&\!\!-1.8088916$\cdot{10^{-2}}$
				\!\!&\!\!-1.9921591$\cdot{10^{-4}}$
				\!\!&\!\!-1.6422068$\cdot{10^{-5}}$       
				\\
				%
				$4d_{5/2}$
				\!\!&\!\!$-3$ 
				\!\!&\!\!-2.4729857$\cdot{10^{-4}}$
				\!\!&\!\!-3.4598067$\cdot{10^{-6}}$
				\!\!&\!\!-3.8498576$\cdot{10^{-7}}$           
				\!\!&\!\!-1.8088823$\cdot{10^{-2}}$
				\!\!&\!\!-1.9921508$\cdot{10^{-4}}$
				\!\!&\!\!-1.6422045$\cdot{10^{-5}}$       
				\\
				%
				$4f_{5/2}$
				\!\!&\!\!$+3$ 
				\!\!&\!\!-4.5984890$\cdot{10^{-6}}$
				\!\!&\!\!-6.4383418$\cdot{10^{-8}}$
				\!\!&\!\!-9.7052084$\cdot{10^{-9}}$            
				\!\!&\!\!-1.8361008$\cdot{10^{-3}}$
				\!\!&\!\!-2.6515938$\cdot{10^{-5}}$
				\!\!&\!\!-3.0478845$\cdot{10^{-6}}$       
				\\
				%
				$4f_{7/2}$
				\!\!&\!\!$-4$ 
				\!\!&\!\!-4.5984835$\cdot{10^{-6}}$
				\!\!&\!\!-6.4383351$\cdot{10^{-8}}$
				\!\!&\!\!-9.7051988$\cdot{10^{-6}}$            
				\!\!&\!\!-1.8360956$\cdot{10^{-3}}$
				\!\!&\!\!-2.6515873$\cdot{10^{-5}}$
				\!\!&\!\!-3.0478779$\cdot{10^{-6}}$       
				\\
				\hline
				$5s_{1/2}$
				\!\!&\!\!$-1$ 
				\!\!&\!\!-2.3068171$\cdot{10^{-2}}$
				\!\!&\!\!-2.7904635$\cdot{10^{-4}}$
				\!\!&\!\! 3.0758120$\cdot{10^{-5}}$            
				\!\!&\!\!-1.4813228$\cdot{10^{-1}}$
				\!\!&\!\!-7.7410848$\cdot{10^{-4}}$
				\!\!&\!\! 3.2499175$\cdot{10^{-4}}$      
				\\
				%
				$5p_{1/2}$
				\!\!&\!\!$+1$ 
				\!\!&\!\!-2.5443058$\cdot{10^{-3}}$
				\!\!&\!\!-5.0167795$\cdot{10^{-5}}$
				\!\!&\!\!-1.5648296$\cdot{10^{-6}}$            
				\!\!&\!\!-4.5359931$\cdot{10^{-2}}$
				\!\!&\!\!-3.5741716$\cdot{10^{-4}}$
				\!\!&\!\! 9.9266091$\cdot{10^{-6}}$       
				\\
				%
				$5p_{3/2}$
				\!\!&\!\!$-2$ 
				\!\!&\!\!-2.5442943$\cdot{10^{-3}}$
				\!\!&\!\!-5.0167631$\cdot{10^{-5}}$
				\!\!&\!\!-1.5648367$\cdot{10^{-5}}$            
				\!\!&\!\!-4.5359271$\cdot{10^{-2}}$
				\!\!&\!\!-3.5741336$\cdot{10^{-4}}$
				\!\!&\!\! 9.9255934$\cdot{10^{-6}}$       
				\\
				%
				$5d_{3/2}$
				\!\!&\!\!$+2$ 
				\!\!&\!\!-2.4729903$\cdot{10^{-4}}$
				\!\!&\!\!-3.4598119$\cdot{10^{-6}}$
				\!\!&\!\!-2.1636864$\cdot{10^{-7}}$            
				\!\!&\!\!-9.6488853$\cdot{10^{-3}}$
				\!\!&\!\!-1.0462533$\cdot{10^{-4}}$
				\!\!&\!\!-8.3402250$\cdot{10^{-6}}$       
				\\
				%
				$5d_{5/2}$
				\!\!&\!\!$-3$ 
				\!\!&\!\!-2.4729857$\cdot{10^{-4}}$
				\!\!&\!\!-3.4598067$\cdot{10^{-6}}$
				\!\!&\!\!-2.1636840$\cdot{10^{-7}}$           
				\!\!&\!\!-9.6488355$\cdot{10^{-3}}$
				\!\!&\!\!-1.0462489$\cdot{10^{-4}}$
				\!\!&\!\!-8.3402159$\cdot{10^{-6}}$       
				\\
				%
				$5f_{5/2}$
				\!\!&\!\!$+3$ 
				\!\!&\!\!-4.5984890$\cdot{10^{-6}}$
				\!\!&\!\!-8.0260480$\cdot{10^{-8}}$
				\!\!&\!\!-7.7854603$\cdot{10^{-9}}$            
				\!\!&\!\!-1.2641487$\cdot{10^{-3}}$
				\!\!&\!\!-1.7661783$\cdot{10^{-5}}$
				\!\!&\!\!-1.9992710$\cdot{10^{-6}}$       
				\\
				%
				$5f_{7/2}$
				\!\!&\!\!$-4$ 
				\!\!&\!\!-4.5984835$\cdot{10^{-6}}$
				\!\!&\!\!-8.0260397$\cdot{10^{-8}}$
				\!\!&\!\!-7.7854526$\cdot{10^{-9}}$           
				\!\!&\!\!-1.2641450$\cdot{10^{-3}}$
				\!\!&\!\!-1.7661739$\cdot{10^{-5}}$
				\!\!&\!\!-1.9992667$\cdot{10^{-6}}$       
				\\
				$5g_{7/2}$
				\!\!&\!\!$+4$ 
				\!\!&\!\!-3.7750656$\cdot{10^{-8}}$
				\!\!&\!\!-0.7700092$\cdot{10^{-9}}$
				\!\!&\!\!-0.0927066$\cdot{10^{-9}}$            
				\!\!&\!\!-7.6482739$\cdot{10^{-5}}$
				\!\!&\!\!-1.3143021$\cdot{10^{-6}}$
				\!\!&\!\!-1.5986779$\cdot{10^{-7}}$       
				\\
				$5g_{9/2}$
				\!\!&\!\!$-5$ 
				\!\!&\!\!-3.7750621$\cdot{10^{-8}}$
				\!\!&\!\!-0.7700085$\cdot{10^{-9}}$
				\!\!&\!\!-0.0927065$\cdot{10^{-9}}$            
				\!\!&\!\!-7.6482586$\cdot{10^{-5}}$
				\!\!&\!\!-1.3142998$\cdot{10^{-6}}$
				\!\!&\!\!-1.5986751$\cdot{10^{-7}}$       
				\\
				\hline 
			\end{tabular}
		\end{small}
		\caption{QED corrections to the  $\,({H}^+\!,\,\pi^-)\,$ 
			and $\,({H}^+\!,\,K^-)\,$
			levels.
		In the first two columns we indicate the state and the 
		corresponding $\kappa$. In the
	last six columns of the first line we specify the atomic systems 
	and the potentials responsible for the energy corrections. 
	The results are expressed in eV.}
	\end{center}
\end{table}
%

\section*{Appendix III. $~$How the reduced mass can be brought to bear in a two-body relativistic framework.}
\bigskip

\renewcommand{\theequation}{III.\arabic{equation}}
The notion of `reduced mass' comes from the
reduction of the Newtonian two-body dynamics and is indeed 
artificial in a relativistic formulation of a two-body problem. 
Despite this, however, 
this quantity may produce some  benefit even in a relativistic 
treatment, giving an approximation of the levels to the order $\alpha$.
The accuracy could possibly  be increased by
carrying on series expansions to higher orders.
Indeed we are able to single out the interaction
term of (\ref{sysdirac}) leading to the reduced mass 
and we give some arguments to show the connection.

We recall that the levels of
the Dirac equation in a Coulomb field with coupling constant $\alpha$, in units of the fermion mass have the well known analytic expression
\begin{eqnarray}
\epsilon_D(\alpha, \kappa, n_r) = \biggl(\,1+\frac{\alpha^2}{\Bigl(\,n_r+\sqrt{\kappa^2-\alpha^2}\,\Bigr)^2}\,\biggr)^{-1/2} -1
\label{diraclevels}
\end{eqnarray}
where $\,n_r\,$ is the radial quantum number \cite{BLP}. For future convenience we observe  that,
for constant $c$,
\begin{eqnarray}
\epsilon_D(c\,\alpha, \kappa, n_r) = 
c^2\,\epsilon_D(\alpha, \kappa, n_r) + O(\alpha^4)
\label{scaladirac}
\end{eqnarray}
Taking into account the definitions of the parameters, 
if we develop the two equations (\ref{sysdirac}) to the
first order in $\alpha$ and the subsequent result 
to the first order in $\epsilon$ we find a new
Dirac equation in the Coulomb field with interaction
constant $\mu_R\,\alpha$ and eigenvalue $\mu_R\,\epsilon$, namely
\begin{eqnarray}
&{}&\frac{d}{ds}\phi(s)+\frac{\kappa}{s}\,\phi(s) -
\Bigl(\,\mu_R\,\frac{\alpha}s+\mu_R\,\epsilon+2\,\Bigr)
\,\chi(s)=0\spazio{1.0}\cr
&{}&\frac{d}{ds}\chi(s)-
\frac{\kappa}{s}\,\,\chi(s) + \Bigl(\,\mu_R\,\frac{\alpha}s+\mu_R\,\epsilon\,\Bigr)\,
\phi(s) =0
\label{sysdiracdev}
\end{eqnarray}
where $\mu_R=\sigma/(1+\sigma)$. Therefore, using
the relation (\ref{scaladirac}) with $c=\mu_R$, the solution 
to the first order in $\alpha$ of the original system (\ref{sysdirac})
is given by
\begin{eqnarray}
\epsilon = 
\mu_R\,\epsilon_D(\alpha, \kappa, n_r)
\label{scaladiracmu}
\end{eqnarray}
Multiplying both sides by the fermion mass $m_F$, we
finally see that 
\begin{equation} 
E=m_F\,\epsilon = m_R\,\epsilon_D(\alpha, \kappa, n_r)
\end{equation}
where $m_R =m_S\,m_F/(m_S+m_F)$ is the classical reduced mass. 
Notice that for $\sigma =1$, namely when the scalar and fermion 
masses are equal, the coefficient $B$ 
is vanishing and (\ref{sysdirac}) is 
exactly the Dirac equation (\ref{sysdiracdev})
with $\mu_R=1/2$. In this case the Coulomb 
coupling constant is $\alpha/2$ and
$\Lambda=\lambda/2$.
The  perturbation treatment of (\ref{sysdirac})
could start from the eigenvalues (\ref{diraclevels})
with $\alpha$ substituted by $\mu_R\,\alpha$. The corresponding
eigenfunctions of (\ref{sysdiracdev})
\begin{eqnarray} 
&{}&\phi =\phantom{-} e^{\displaystyle{-as}}\,(as)^{\displaystyle{-1+\gamma}}\,(\,1+\mu_R\epsilon\,)^{1/2}\,
\Bigl(\,-N_1\,L_{n_r-1}^{2\gamma}(2as) + N_2\,L_{n_r}^{2\gamma}(2as)\,\Bigr )
\spazio{1.0}\cr
&{}&\chi =-e^{\displaystyle{-as}}\,(as)^{\displaystyle{-1+\gamma}}\,(\,1-\mu_R\epsilon\,)^{1/2}\,
\Bigl(\,\phantom{-}N_1\,L_{n_r-1}^{2\gamma}(2as) + N_2\,L_{n_r}^{2\gamma}(2as)\,\Bigr )
\end{eqnarray}
where $L_n^b$ are the associated Legendre polynomials \cite{Pid},
$\gamma=(\kappa^2+\mu_R^2\alpha^2)^{1/2}$,
$a=\mu_R\alpha\,(\kappa^2+n_r^2+2n_r\gamma)^{-1/2}$
and the normalization constants $N_1,\,N_2$ can be recovered
from \cite{BLP}. We finally observe that, being a genuine 
Dirac equation, (\ref{sysdiracdev}) conserves the
Johnson-Lippmann operator which is not conserved in (\ref{sysdirac}). 
This suggests that indeed the 
breaking has to be very small, as confirmed by the
numerical results.


\begin{thebibliography}{999}
\bigskip


\bibitem{GS1} R. Giachetti, E. Sorace, {J. Phys. A}, {\bf 38}, 1345 (2005).

\bibitem{GS2} R. Giachetti, E. Sorace, {J. Phys. A} {\bf 39}, 15207 (2006).

\bibitem{GS3} R. Giachetti, E. Sorace, {Phys. Rev. D}, {\bf 87}, 034021 (2013).

\bibitem{BGS_JPB} A. Barducci, R. Giachetti, E. Sorace, {J. Phys. B: At. Mol. Opt. Phys.} {\bf 48},
085002 (2015).

\bibitem{BGS_PR}  A. Barducci, R. Giachetti, E. Sorace,  {Phys. Rev. D}, {\bf 95}, 054022 (2017).

\bibitem{GS-AP} R. Giachetti, E. Sorace, {Annals of Physics} {\bf 401}, 202 (2019). 

\bibitem{Iwa} M. Iwasaki et al., {Nucl. Phys. A}
 {\bf 639}, 501 (1998). 

\bibitem{Schroeder} H.Ch. Schr\"oder et al., {Phys. Lett. B} {\bf 469}, 25 (1999).

\bibitem{Curc} C. Curceanu, C. Guaraldo et al., 
{Symmetry}, {\bf 12}, 547 (2020).

\bibitem{Bazzi} M. Bazzi et al.,  {Phys. Lett. B} {\bf 704}, 113 (2011).

\bibitem{Matsinos} E. Matsinos, {\it{A brief history of the pion-nucleon coupling constant}},
e-Print: arXiv:1901.01204 (2019).

\bibitem{IndeTra} M. Trassinelli, P. Indelicato,  {Phys. Rev. A} {\bf 76 }, 012510 (2007).

\bibitem{Tra} M. Trassinelli,  {Phys. Lett. B} {\bf 759}, 583 (2016).

\bibitem{SBIP} S. Schlesser, E.O. Le Bigot, P.
Indelicato, K. Pachucki, {Phys. Rev. C} {\bf 84 },
015211 (2011).

\bibitem{Guy} G. Ron, {\it{The proton radius}},
Talk 11/01/2018, Open Access at Springerlink.com,\\
https://www.npl.washington.edu/sites/default/files/speaker-materials/Proton

\bibitem{Pato} A.S.M. Patoary, N.S. Oreshkina,
{Eur. Phys. J.} {\bf D72 }, 54 (2018).

\bibitem{XGG} W. Xiong, A. Gasparian, H.Gao et al.,
{Nature} {\bf 575 }, 147 (2019).

\bibitem{DFK} M. Daum, R. Frosch, P.R. Kettle,
{Phys. Lett. B} {\bf 796}, 11 (2019).

\bibitem{BLP} V.B. Berestetskii, E.M. Lifshitz, 
L.P. Pitaevskii,\textit{ Quantum 
	Electrodynamics} (Pergamon Press, 1982).

\bibitem{MF} A.P. Martynenko, R.N. Faustov,
{Theor. Math. Phys.} {\bf 64}, 765 (1985).

\bibitem{FV} H. Feshbach, F. Villars, Rev. Mod. Phys. 30 (1958) 1.

\bibitem{Cope} F.T. Cope, Am. J. Math. {\bf 56}, 411
(1934) and {\bf 58} 130 (1936).

\bibitem{PDG} \textit{Review of Particle Physics}, Particle Data Group, Chinese Physics C, \textbf{38}, Number 9, (2014).

\bibitem{BS} R. Beals, J. Szmigielski, {Notices of the AMS} {\bf 60}, 866 (2013),

\bibitem{Hetal} M. Hennebach et al.,  Eur. Phys. J. {\bf A50}, 329 (2014), Erratum: Eur .Phys. J. {\bf A55}, 24 (2019).

\bibitem{Getal} D. Gotta et al., Physics Procedia {\bf 17}, 69 (2011).
    
\bibitem{cata} C. Corceanu et al., Rev. Mod Phys. 
{91}, 1 (2019).

\bibitem{iwa} M. Iwasaki et al., Phys. Rev. Lett.
{\bf 78}, 3067 (1997).

\bibitem{JL} M.H. Johnson, B.A. Lippmann, 
{Phys. Rev} {\bf 78 }, 329 (1950).

\bibitem{Marton} J. Marton et al., EPJ Web of Conference  {\bf 199}, 03004 (2019).

\bibitem{Pid} F.B. Pidduck, London J. of London Math. Soc. {\bf s1-4}, 163 (1929).





\end{thebibliography}
\end{document}